\documentclass[aps,prc,twocolumn,nofootinbib,groupedaddress,showpacs]{revtex4}
\usepackage{epsfig}
\usepackage{dcolumn}

\def\bea {\begin{eqnarray}}
\def\eea {\end{eqnarray}}
\def\be {\begin{equation}}
\def\ee {\end{equation}}
\def\ben{\begin{enumerate}}
\def\een{\end{enumerate}}
\def\bi{\begin{itemize}}
\def\ei{\end{itemize}}

\def\F{{\cal F}}

\def\GV{G_{\mbox{\tiny V}}}
\def\GA{G_{\mbox{\tiny A}}}

\def\DRV{\Delta_{\mbox{\tiny R}}^{\mbox{\tiny V}}}
\def\DRA{\Delta_{\mbox{\tiny R}}^{\mbox{\tiny A}}}
\def\fV{f_{\mbox{\tiny V}}}
\def\fA{f_{\mbox{\tiny A}}}

\def\hyphen{{\mbox{-}}}

\def\2p{|2p\rangle }
\def\4p2h{|4p\hyphen 2h\rangle }
\def\6p4h{|6p\hyphen 4h\rangle }

\begin{document} 
\title{Branching Ratios for the Beta Decay of $^{21}$Na}
\author{V.E. Iacob, J.C. Hardy, C.A. Gagliardi, J. Goodwin, N. Nica, H.I. Park, G. Tabacaru, L. Trache, R.E. Tribble,
Y. Zhai and I.S. Towner\footnote{Present address: Physics Department, Queen's University, Kingston,
Ontario K7L 3N6, Canada}}
\affiliation{Cyclotron Institute, Texas A \& M University, College Station, Texas 77843}                    

\date{\today} 
\begin{abstract} 

We have measured the beta-decay branching ratio for the transition from $^{21}$Na to the
first excited state of $^{21}$Ne.  A recently published test of the standard model, which
was based on a measurement of the $\beta$-$\nu$ correlation in the decay of $^{21}$Na, 
depended on this branching ratio.  However, until now only relatively imprecise
(and, in some cases, contradictory) values existed for it.  Our new result, $4.74(4)\%$,
reduces but does not remove the reported discrepancy with the standard model.
\end{abstract} 


\maketitle

\section{Introduction}
\label{intro}

A recent publication by Scielzo {\it et al.}\cite{Sc04} reported a measurement of the
$\beta$-$\nu$ angular correlation coefficient, $a_{\beta\nu}$, for the $\beta$-decay
transition between $^{21}$Na and the ground state of its mirror, $^{21}$Ne.  
The authors compare their result with the standard-model prediction for $a_{\beta\nu}$,
with a view to testing for scalar or tensor currents, the presence of which would
signal the need for an extension of the standard model.  Although they found a
significant discrepancy -- the measured value, $a_{\beta\nu} = 0.524(9)$, disagrees
with the standard-model prediction of 0.558 -- they stop short of claiming a
fundamental disagreement with the standard model. 

\begin{figure}[b]
\epsfig{file=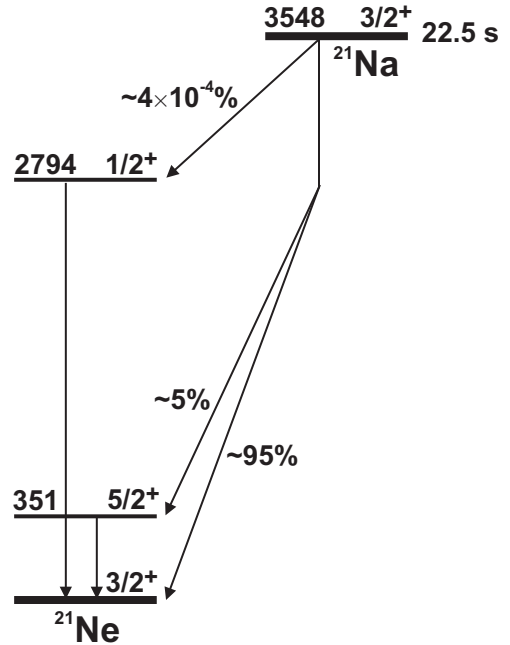,width=6.5cm}
\caption{$\beta$-decay scheme for $^{21}$Na}
\label{fig:1}
\end{figure}

Scielzo {\it et al.} \cite{Sc04} offer two alternative explanations that would have to
be eliminated before their result could begin to raise questions about the need for an
extension to the standard model.  One is that some $^{21}$Na$_2$ dimers formed by cold
photoassociation could also have been present in their trap, thus distorting the result;
they themselves propose to do further measurements to test that possibility.  The other
is that the branching-ratio value they used for $\beta$ decay to the first excited state of
$^{21}$Ne might not be correct.  Because Scielzo's measurement could not distinguish between
positrons from the two predominant $\beta$-decay branches from $^{21}$Na (see Fig. \ref{fig:1}),
the adopted branching ratio for the $\beta$ transition to the first excited state not only affects 
their data analysis but also helps determine the theoretical prediction for $a_{\beta\nu}$
itself, since the axial-vector component of the ground-state branch can only be determined
from its $ft$ value, which also depends on the branching ratio.  This branching ratio is a
key component of their standard-model test, yet the five published values \cite{Ta60,Ar63,Al74,Az77,Wi80}
are between 25 and 45 years old, are quite inconsistent with one another and range from
2.2(3) to 5.1(2)\%.  To remedy this problem, we report here a new measurement of the ground-state
branching ratio, for which we quote $\pm$0.8\% relative precision, five times better than the best
precision claimed in any previous measurement.  

\section{Experiment}
\label{exp}

We produced 22.5-s $^{21}$Na using a 28{\it A}-MeV $^{22}$Ne beam from the Texas A\&M K500
superconducting cyclotron to initiate the $^1$H($^{22}$Ne, $2n$)$^{21}$Na reaction on a
LN$_2$-cooled hydrogen gas target.  The ejectiles from the reaction were fully stripped and, after
passing through the MARS spectrometer \cite{Tr91}, produced a $^{21}$Na secondary beam of $>$99\%
purity at the extraction slits in the MARS focal plane.  This beam, containing $\sim$$3\times10^5$
atoms/s at $24.4A$ MeV, then exited the vacuum system through a 50-$\mu$m-thick Kapton window,
passed successively through a 0.3-mm-thick BC-404 scintillator and a stack of aluminum
degraders, finally stopping in the 76-$\mu$m-thick aluminized mylar tape of a tape transport
system.  Since the few impurities remaining in the beam had ranges different from that of $^{21}$Na, 
most were not collected on the tape; residual collected impurities were concluded to be less
than $0.1$\% of the $^{21}$Na content.

\begin{figure}[t]
\epsfig{file=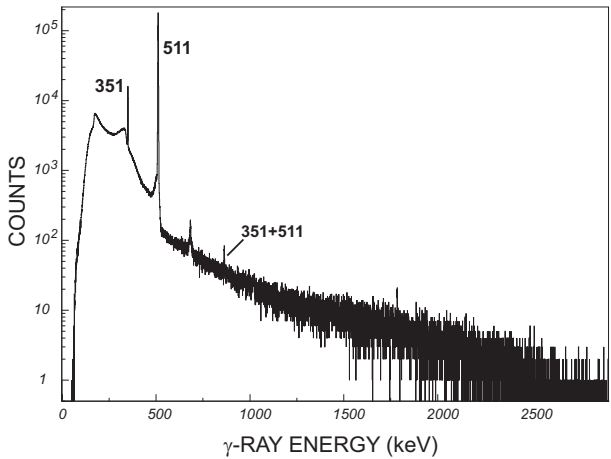,width=8cm}
\caption{Spectrum of $\beta$-delayed $\gamma$ rays observed in coincidence with positrons following
the decay of $^{21}$Na.  It includes about half of the
total data collected.  The peaks attributable to $^{21}$Na are marked with their energy in keV;
the sum peak is identified by its components.  The small unmarked peak at 682 keV is caused by
summing of one 511-keV $\gamma$ ray with the back-scattered $\gamma$ ray (171 keV) from the
second 511-keV $\gamma$ ray.}
\label{fig:2}
\end{figure}
 
In a typical measurement, we collected $^{21}$Na on the tape for a few seconds, then
interrupted the beam and triggered the tape-transport system to move the sample in 180 ms to
a shielded counting station located 90 cm away, where the sample was positioned between
a 1-mm-thick BC404 scintillator to detect $\beta^+$ particles, and a 70\% HPGe detector for
$\gamma$ rays.  Two timing modes were used: in one, the collection and detection periods
were 3 and 30 s, respectively; in the other, they were 6 and 60 s.  In both cases, after the
detection period was complete, the cycle was repeated and, in all, some 3,200 cycles were
completed over a span of 32 hours.

Time-tagged $\beta$-$\gamma$ coincidence data were stored event by event.  The $\beta$ and
$\gamma$-ray energies, the coincidence time between them, and the time of the event after
the beginning of the cycle were all recorded, as was the total number of $\beta$-singles
events for each cycle.  The same discriminator signal used for scaling was also used in
establishing the $\beta$-$\gamma$ coincidences.

Essential to our experimental method is the precise absolute efficiency of the $\gamma$-ray
detector, which was positioned 15 cm from the collected sample.  We have meticulously
calibrated our HPGe detector at this distance over a five-year period using, 
in total, 13 individual sources from 10 different radionuclides: $^{48}$Cr, $^{60}$Co, 
$^{88}$Y, $^{108m}$Ag, $^{109}$Cd, $^{120m}$Sb, $^{133}$Ba, $^{134}$Cs, $^{137}$Cs and
$^{180m}$Hf.  Two of the $^{60}$Co sources were specially prepared by the
Physikalisch-Technische Bundesanstalt \cite{Sc02} with activities certified to 0.06\%.  The
details of our calibration procedures, which include both source measurements and Monte
Carlo calculations, have been published elsewhere \cite{Ha02,He03,He04}.  The absolute
efficiency of our detector is known to 0.2\% in the energy range from 50 to 1400 keV, and
to 0.4\% from 1400 keV to 3.5 MeV.

The absolute efficiency of the $\beta^+$ detector, which was located 1.5 cm from the collected
sample, is not required for our measurement but its dependence on energy is of some importance
(see section \ref{res}).  We have explored the efficiency of this detector {\it via} measurements
and Monte Carlo calculations, and its dependence on $\beta^+$ energy is now reasonably well
understood \cite{Ia06}.

A typical $\gamma$-ray spectrum recorded in coincidence with betas is presented in Fig. \ref{fig:2}.
Apart from the annihilation radiation, the only significant peak in the spectrum is the 351-keV
$\gamma$ ray from the first excited state in $^{21}$Ne.  In 3,200 total cycles we recorded more
than $8\times10^4$ counts in this peak.

It was important to our later analysis that we establish the contribution of room background both
to the $\beta$-$\gamma$ coincidence spectrum and to the $\beta$-detector singles rate.  For this
purpose, we recorded data with the cyclotron beam on but with a thick degrader inserted just
upstream from the tape; everything was thus identical to a normal measurement except that no
$^{21}$Na was implanted in the tape.  Both the coincidence and singles rates were observed to
drop to 0.04\% of the rate observed when $^{21}$Na was correctly implanted.  Room background
was thus effectively negligible in our analysis.

\section{Results}
\label{res}

The $\beta$-decay scheme of $^{21}$Na is shown in Fig. \ref{fig:1}.  The only branch in addition
to those populating the ground and first excited states is known to be very weak \cite{Wi80}
and can be ignored in our analysis.  In that case, the branching ratio, $R_1$, for population
of the first excited state can be determined from the measured intensity ratio of the 351-keV $\gamma$  
ray relative to the total number of $^{21}$Na decays.  Thus, we obtain $R_1$ from the following
relationship:
\be
R_1 = \frac{N_{\gamma \beta}}{N_{\beta} \epsilon_\gamma} k ,
\label{R}
\ee
\noindent where $N_{\gamma \beta}$ is the number of 351-keV $\gamma$ rays observed in coincidence
with betas; $N_{\beta}$ is the number of (singles) betas observed; $\epsilon_{\gamma}$ is the
efficiency of the HPGe detector for 351-keV $\gamma$ rays; and $k$ is a factor ($\sim$ 1) that
accounts for small experimental corrections that will be enumerated in what follows.  Note that
the efficiency of the beta detector does not appear in Eq. \ref{R}, although its dependence on
$\beta^+$ energy will be seen to play a minor role in the evaluation of $k$.

Before determining the ratio $N_{\gamma \beta}/N_{\beta}$ from our data, we eliminated those
cycles in which the collected source was not positioned exactly between the $\beta$ and $\gamma$
detectors.  Although the tape-transport system is quite consistent in placing the collected
source within $\pm3$ mm of the designated counting location, it is a mechanical system, and
occasionally larger deviations occur.  For each cycle we recorded not only the total number of
positrons detected but also the total number of $^{21}$Na ions that emerged from the MARS
spectrometer, as detected by the scintillator located immediately in front of the aluminum
degraders.  The ratio of the former to the latter is a very sensitive measure of how well the
source is positioned with respect to the $\beta$ detector.  In analyzing the data, we rejected
the results from any cycle with an anomalous (low) ratio.  Under these conditions, we obtained
the result $N_{\gamma \beta}/N_{\beta} = 2.378(13)\times 10^{-4}$.

As stated in section \ref{exp}, the absolute efficiency, $\epsilon_\gamma$, of our detector at
15 cm is known to $\pm 0.2\%$.  However, this applies to a highly controlled situation in which
the source-to-detector distance can be measured by micrometer to a small fraction of a millimeter.  
With the fast tape-transport delivery system, we cannot be assured of reproducibility at the same
level of precision.  Taking $\pm 0.5$ mm to be our actual uncertainty in position under experimental
conditions, we add an uncertainty of $\pm 0.6\%$ to the detector efficiency in quadrature with the
basic $\pm 0.2\%$ uncertainty.  For the 351-keV $\gamma$ ray, this leads to $\epsilon_\gamma = 
5.12(3)\times 10^{-3}$, the value we insert in Eq. \ref{R}.

Although the ratio $N_{\gamma \beta}/N_{\beta}$ and $\epsilon_\gamma$ are the predominant
experimental quantities required to evaluate the branching ratio, it is the correction factor $k$
that holds the key to our achieving high precision.  In fact, $k$ is really a product of four
separate corrections, $k_1 ... k_4$.  We will deal with each individually.

{\it Random coincidences $(k_1)$. ---} Since the time between each coincident $\beta$ and $\gamma$ ray
was recorded event by event, we could project out the time spectrum corresponding to the 351-keV $\gamma$
ray.  In that spectrum, the prompt coincidence peak stood prominently above the flat random distribution, 
allowing us clearly to distinguish the relative contributions of real and random coincidences.  The correction
factor required to account for the random contribution to the $\beta$-coincident 351-keV $\gamma$-ray
peak was thus determined to be $k_1$ = 0.9884(10).  Naturally, this correction accounts not only for random
coincidences among $^{21}$Na $\beta$ and $\gamma$ rays but also for random coincidences between $^{21}$Na
betas and any $\gamma$ rays originating from room background. 

{\it Real-coincidence summing $(k_2)$.---} Since each 351-keV $\gamma$ ray from the decay of the first excited
state in $^{21}$Ne is accompanied by a positron from the $^{21}$Na $\beta^+$-decay branch that populated
the state, there is a significant probability that a 351-keV $\gamma$ ray and 511-keV annihilation
radiation will reach our HPGe detector simultaneously and be recorded as a single $\gamma$ ray with
the combined energy of both.  Any summing of this kind will rob events from the 351-keV photopeak.  
Our first step in accounting for the resultant loss was to obtain the area of the observed 862-keV
(511+351) sum peak.  Since losses from the 351-keV photopeak result not just from its summing with the
511-keV photopeak but also with the latter's Compton scattered radiation, as a second step we multiplied the
sum-peak area by the known ``total-to-peak" ratio for our detector at 511 keV (see Fig. 11 in
reference \cite{He03}).  Finally, this result for losses was increased by 4\% to account for
annihilation in flight, which leads to 351-keV peak summing with annihilation radiation of different
energies, and by another 2.5\% to account for summing with positrons backscattered from the plastic
scintillator.  The total loss due to real-coincidence summing was thus determined to be 1.78\%: {\it i.e.}
$k_2$ = 1.0178(17).

{\it Dead time $(k_3)$.---} Equation \ref{R} depends upon $N_{\gamma \beta}$ and $N_{\beta}$ being recorded
for identical times.  In our experiment they were, of course, gated on and off together, but during the
counting period the circuit dead time for $N_{\gamma \beta}$, which was limited by the relatively slow
electronics used for $\gamma$-ray counting, was much greater than that for $N_{\beta}$, which was simply
scaled.  We determined the dead time associated with $N_{\gamma \beta}$ from the total rate in the
HPGe detector during the counting period and from the known processing time (32 $\mu$s) for each coincident
event.  The scaler dead time per event was only 100 ns but the total rate in the scaler was much higher
than the HPGe rate; nevertheless the dead time associated with $N_{\beta}$ turned out to be smaller by a
factor of three than that associated with coincidence events.  The overall correction factor is
$k_3$ = 1.0018(1).

\begin{table} [b]
\begin{center}
\caption{Error budget for the measured branching ratio $R_1$ 
\label{err}}
\begin{ruledtabular}
\begin{tabular}{lc}
&  \\ [-3mm]
\multicolumn{1}{l}{Origin of uncertainty}
& \multicolumn{1}{c}{$\%$ uncertainty} \\
& \\ [-3mm]
\hline
&  \\ [-3mm]
Experimental ratio, $N_{\gamma \beta}/N_{\beta}$ & 0.52 \\
HPGe detector efficiency & 0.20 \\
Source-detector distance & 0.60 \\
Random coincidences & 0.11 \\
Real-coincidence summing & 0.17 \\
Dead time & 0.01 \\
$\beta$-detector efficiency {\it vs} energy & 0.13 \\
&  \\
Total uncertainty on $R_1$ & 0.85 \\
\end{tabular}
\end{ruledtabular}
\vspace{-0.5cm}
\end{center}
\end{table}

{\it Beta-detector response function $(k_4)$.---} The correction factor associated with the $\beta$-detector
response function is given by
\be
k_4 = \frac{\epsilon_{\beta_{total}}}{\epsilon_{\beta_1}}
    \simeq \frac{0.95\epsilon_{\beta_0}+0.05\epsilon_{\beta_1}}{\epsilon_{\beta_1}} ,
\label{k}
\ee
\noindent where $\epsilon_{\beta_0}$ and $\epsilon_{\beta_1}$ are the detector efficiencies for the
$\beta$ transitions to the ground and first excited states respectively.  If the detector response
function were completely independent of energy, then this correction factor would be unity.  In fact,
though, the efficiency does change slightly with energy.  We have studied this effect using
measurements with sources -- $^{90}$Sr, $^{133}$Ba, $^{137}$Cs and $^{207}$Bi -- aided by Monte
Carlo calculations \cite{Ia06}.  Including the effects of our low-energy electronic threshold, we
determine that $\epsilon_{\beta_1}/\epsilon_{\beta_0}$ = 0.987, which leads to the correction factor
$k_4$ = 1.0129(13).

Multiplying $k_1$ through $k_4$ we determine the correction factor in Eq. \ref{R} to be $k$ = 1.0208(24).  
When combined in Eq. \ref{R} with the other factors already discussed, this yields the final result for
the branching ratio to the first excited state in $^{21}$Na:
\be
R_1 = 0.0474(4).
\label{Result}
\ee
\noindent The complete error budget corresponding to our quoted $\pm0.85\%$ uncertainty is given in Table \ref{err}.

\section{Analysis}
\label{ana}

Our measured branching-ratio value is compared with previous measurements in Table \ref{pre}.  All
previous experiments determined the branching ratio from a comparison of the area of the 351-keV
peak to that of the annihilation radiation.  This method has the advantage that only relative
detector efficiencies are required, but it has three serious disadvantages: i) contaminant activities
may well make an unknown contribution to the annihilation radiation; ii) most positrons do not annihilate
at the source position, where the $\gamma$ rays originate, so the relative detection efficiencies
cannot be simply determined from calibration sources; and iii) the significant effect ($\sim$5\%)
of positron annihilation in flight is a first-order correction that must be calculated and corrected
for.  All previous measurements except possibly reference \cite{Al74} were susceptible to potential
contaminants; only the last three references \cite{Al74,Az77,Wi80} mention accounting for a spatially
distributed source of 511-keV radiation; and only the last two \cite{Az77,Wi80} appear to have taken
account of annihilation in flight.

\begin{table} [b]
\begin{center}
\caption{Measurements of the branching ratio $R_1$ 
\label{pre}}
\begin{ruledtabular}
\begin{tabular}{lll}
& & \\ [-3mm]
\multicolumn{1}{l}{Date}
& \multicolumn{1}{l}{Reference}
& \multicolumn{1}{l}{Result(\%)} \\
& & \\ [-3mm]
\hline
& & \\ [-3mm]
1960 & Talbert \& Stewart \cite{Ta60} & 2.2(3) \\
1963 & Arnell \& Wernbom \cite{Ar63} & 2.3(2) \\
1974 & Alburger \cite{Al74} & 5.1(2) \\
1977 & Azuelos, Kitching \& Ramavataram \cite{Az77} & 4.2(2) \\
1980 & Wilson, Kavanagh \& Mann \cite{Wi80} & 4.97(16) \\
 & & \\
2006 & This measurement & 4.74(4) \\
\end{tabular}
\end{ruledtabular}
\vspace{-0.5cm}
\end{center}
\end{table}

Given the age of the previous measurements and the potential hazards associated with their experimental
method -- not to mention their mutual inconsistency -- we choose not to average our result with them
but instead to use our present result alone in extracting the properties of the $^{21}$Na $\beta$-decay
scheme.

Since there are only two significant $\beta$-decay branches from $^{21}$Na -- to the ground and first excited
states of the daughter -- with $R_1$ determined, the branching ratio to the ground state, $R_0$, follows
directly from it:
\be
R_0 = 0.9526(4) ,
\label{R0}
\ee    
\noindent where this result is actually determined to a precision of 0.04\%.  We now proceed from this value
for $R_0$ to obtain the $ft$ value for this transition, the relative contributions of axial-vector and vector
components, and ultimately the standard-model expectation for its $\beta$-$\nu$ angular correlation coefficient.

In deriving the $ft$ value for the ground-state mirror transition, we take the half-life of $^{21}$Na
to be $t_{1/2}$ = 22.49(4) s and its total decay energy to be $Q_{EC}$ = 3547.6(7) keV.  The former is the
average of two mutually consistent results \cite{Al74,Az77} and the latter is the value quoted in the 2003
Atomic Mass Evaluation \cite{Au03} where it was obtained from a single $^{20}$Ne(p,$\gamma$)$^{21}$Na
measurement made in 1969 \cite{Bl69} and then revised by Audi {\it et al.} \cite{Au03} to take account of
more up-to-date calibration energies.  With the calculated electron-capture probability for the ground-state
transition being 0.00095, the average half life, when combined with our branching ratio value from Eq. \ref{R0},
yields a partial half-life for the transition of 23.63(4) s.

Next we compute the value of $f$ from the $Q_{EC}$ value following methods similar to those we used in the
analysis of superallowed $\beta$ decay; these are described in the Appendix to reference \cite{Ha05}.  To make an
``exact" calculation that includes, for example, the effects of weak magnetism and other induced corrections we
need a shell-model calculation of the appropriate nuclear matrix elements.  For this we used an $(s,d)$-shell
model space and the universal $(s,d)$-shell effective interaction of Wildenthal \cite{Wi84}.  This interaction
has been demonstrated \cite{Br85} to reproduce energy spectra and Gamow-Teller matrix elements in this mass
region providing that the axial-vector coupling constant is quenched.  In our calculation, we fine-tuned the
amount of quenching to reproduce our experimental data\footnote{We adjusted the quenching so that it reproduced
our measured value of $\lambda$ (see Eq. \ref{lambdaexp}).  This corresponded to $\GA \simeq \GV$ and is
essentially the same result that Brown and Wildenthal \cite{Br85} established for the shell as a whole.}.

For a mirror transition like this one, which includes both vector and axial-vector components, the $f$ value
calculated for the vector part of the weak interaction, $\fV$, is slightly different from the value calculated
for the axial-vector part, $\fA$.  In the allowed approximation it is always assumed that $\fV$ = $\fA$ = $f$ but,
where high precision is sought, a more exact calculation is required.  The results we obtain, $\fV$ = 170.974
and $\fA$ = 174.157, are nearly 2\% different from one another, principally as a result of the influence
of weak magnetism on the shape correction factor of the axial-vector component.  In quoting the $ft$ value
for the mirror ground-state transition, we make the (arbitrary) choice to use $\fV$, with the result that
\be
\fV t = 4040(9) s .
\label{ft}
\ee

Like any other $ft$ value, this result can be related to vector and axial-vector coupling constants, and
to the matrix elements pertaining to the specific transition.  To do so with the precision required for
a standard-model test requires that radiative and charge-dependent corrections be incorporated.  The
expression we use is the following:
\bea
\Bigm{[}\fV\GV^2\langle1\rangle^2(1 + \delta_{NS} - \delta_C )(1 + \DRV)~~~~~~~~~~~ &  &
\nonumber \\
+ \fA\GA^2\langle\sigma\rangle^2(1 - \delta_A)(1
+ \DRA)\Bigm{]}(1 + \delta_R^{\prime})t & = & K ~~~~~~~~
\label{fVfA}
\eea
\noindent where $K/(\hbar c )^6 = ( 8120.271 \pm 0.012 ) \times 10^{-10}$ GeV$^{-4}$s; $\GV$ and
$\GA$ are the vector and axial-vector coupling constants for nuclear weak decay; and $\langle1\rangle$ and
$\langle\sigma\rangle$ are the Fermi (vector) and Gamow-Teller (axial-vector) matrix elements, respectively,
for the ground-state transition.  For this particular transition between T=$\frac{1}{2}$ states, $\langle1\rangle$=1.
The transition-dependent radiative correction terms, $\delta_R^{\prime}$ and $\delta_{NS}$, and the
isospin-symmetry-breaking correction, $\delta_C$, all have their conventional definitions \cite{Ha05} but, 
in the present context of a mixed vector and axial-vector transition, we note that $\delta_R^{\prime}$ is
the same for both components while $\delta_C$ and $\delta_{NS}$ only pertain to the vector component.  The
latter two terms have their equivalents that must be applied to the axial-vector component but we subsume
them into a term we call $\delta_A$: as it turns out, we will not have to calculate a value for $\delta_A$.
Finally, the transition-independent radiative correction also takes on different values for the vector and
axial-vector components, $\DRV$ and $\DRA$; but neither will have to be calculated.

Rearranging Eq. \ref{fVfA}, we obtain the result:
\bea
\fV t(1 + \delta_R^{\prime})(1 + \delta_{NS} - \delta_C ) = 
\frac{K}{\GV^2(1 + \DRV)\bigm{[}1+\lambda^2\frac{\fA}{\fV}\bigm{]}}
\label{ftcor}
\eea
\noindent where
\vspace{-3mm}
\bea 
\lambda = \frac{\GA\langle\sigma\rangle(1 - \delta_A)^{1/2}(1 + \DRA)^{1/2}}{\GV(1 + \delta_{NS} - \delta_C )^{1/2}(1 + \DRV)^{1/2}}.
\nonumber
\label{lambda}
\eea
\noindent Here $\lambda$ is the ratio of axial-vector to vector components in the transition.  A further
simplification in this equation can be achieved by our implementing the results from superallowed
$0^+\rightarrow0^+$ beta decays, which provide an experimental determination of the product $\GV^2(1 + \DRV)$.
The average corrected ${\F t}$ value from these decays \cite{Ha05} is related to the vector coupling
constant via the relationship:
\be
\overline{\F t} = \frac{K}{2\GV^2(1 + \DRV)}.
\label{Ft}
\ee

Since it is the term $\lambda$ that we need to extract from experiment in order to calculate the
$\beta$-$\nu$ angular correlation coefficient, we now re-express Eq. \ref{ftcor} in the following form:
\be
\lambda^2 = \frac{\fV}{\fA}\Bigm{[}\frac{2\overline{\F t}}{\fV t(1 + \delta_R^{\prime})(1 +
\delta_{NS} - \delta_C )}-1 \Bigm{]}.
\label{lambda1}
\ee
We have calculated the three remaining correction terms using the same methods as were described in
reference \cite{To02}, the results being $\delta_R^{\prime}$ = 1.492(15)\%, $\delta_C$ = 0.268(16)\% and
$\delta_{NS}$ = -0.065(20)\%.  We then adopt the value, $\overline{\F t}$ = 3072.7(8), which is the
average result extracted from superallowed $0^+\rightarrow0^+$ beta decays when the correction terms
are calculated by the same methods as those used here (see Eq. 11 in reference \cite{Ha05}).  Thus we
finally obtain
\be
\lambda = 0.7033(24)
\label{lambdaexp}
\ee
\noindent for the ground-state mirror transition.

Based on this result for $\lambda$ we have computed the beta-neutrino correlation coefficient exactly,
following the formalism of Behrens-B\"{u}hring \cite{Be82}.  These authors write the electron-neutrino
correlation $\omega(\theta,W)$ as:
\be
\omega(\theta,W) = \sum_k D(k,W) P_k(\cos \theta) ,
\label{omega}
\ee
\noindent where $W$ is the electron energy (in rest-mass units), $\theta$ is the angle between the
emitted electron and neutrino directions, and $P_k$ are Legendre polynomials.  The sum is over
$k=0,1,2$.  The coefficients $D(k,W)$ are expressed fully by Behrens and B\"{u}hring \cite{Be82},
from which it can be seen that $D(0,W)$ is exactly equal to 1.0, and $D(2,W)$ is small.  The term $D(1,W)$
relates to the beta-neutrino angular correlation coefficient, $a_{\beta\nu}$, {\it via} the
expression
\be
D(1,W) = a_{\beta\nu} p/W ,
\label{D}
\ee
\noindent where $p = \sqrt{W^2-1}$.  For the exact expression for $a_{\beta\nu}$ we compute
\be
a_{\beta\nu} = \langle D(1,W) W/p \rangle ,
\label{a}
\ee
\noindent where $\langle .. \rangle$ signifies an average over the beta spectrum.  It should be noted
that this exact evaluation of $a_{\beta\nu}$ yields a result that is about 1\% different from the
approximate expression that is often used: {\it viz.}
\be
(a_{\beta\nu})_{approx} = (1 - \lambda^2/3)/(1 + \lambda^2) .
\label{approx}
\ee
\noindent The exact expression in Eq. \ref{a} differs from this approximate one by the inclusion of energy
dependence as well as weak magnetism and other small effects.  Our final computed result for the exact
$\beta$-$\nu$ angular correlation coefficient based on our new experimental result for $\lambda$ is
\be
a_{\beta\nu} = 0.553(2) .
\label{exact}
\ee
\noindent This can now stand as the ``standard-model prediction" for $a_{\beta\nu}$, against which the
measured angular-correlation coefficient can be compared.  Our new value is 0.9\% lower than the one
originally used by Scielzo {\it et al.} \cite{Sc04}.

\section{Conclusions}
\label{conc}

As noted in the Introduction, the value of the branching ratio affects not only the standard-model
prediction for $a_{\beta\nu}$ (see Eq. \ref{exact}) but also the analysis by Scielzo {\it et al.}
\cite{Sc04} of their measurement of that coefficient.  With the excited-state branching ratio taken to be
5.02(13)\%, they applied a correction of +6.81(18)\% to their result.  Since this correction scales
with the branching ratio \cite{Sc05}, our new value for the latter leads to a new correction factor
of +6.44(5)\%.  This downward shift of 0.4\% is actually rather small compared to the overall
uncertainties quoted by Scielzo {\it et al.}, and their value for $a_{\beta\nu}$ as obtained from
$^{21}$Ne$^{1+}$ only changes from 0.524(9) to 0.523(9).

As a result of our new measurement we have improved -- and lowered slightly -- the standard-model prediction
of the $\beta$-$\nu$ angular correlation coefficient for the mirror transition from $^{21}$Na.  This new
prediction still leaves the Scielzo {\it et al.} experimental result \cite{Sc04} in disagreement with the
prediction.   However, the authors themselves expressed concern about the possible presence of $^{21}$Na$_2$
dimers in their trapped samples; this would have caused a dependence of their result on the trapped-atom
population and could easily reconcile their result with the standard model.  With a precise branching ratio now
determined, an investigation of the actual make-up of the trapped-atom samples in the Scielzo {\it et al.}
experiment is essential if the $^{21}$Na result is to become a real test of the standard model.

Finally, we draw attention to the fact that the ``standard-model prediction" for $a_{\beta\nu}$ depends
on the half-life of $^{21}$Na and its $\beta$-decay $Q$ value through the $\fV t$ value for the ground-state
transition (see Eq. \ref{ft} and the preceeding paragraphs).  The half-life has only been measured twice
\cite{Al74,Az77} -- in experiments that did not obtain branching ratios in agreement with our current result --
and the $Q$ value comes from a single 35-year-old (p,$\gamma$) measurement \cite{Bl69} originally based on
long-outdated calibration energies.  Clearly, both these results could be improved significantly by modern
measurements.

The authors would like to thank N.D. Scielzo for helpful discussions. This work was supported by the U.S.
Department of Energy under Grant No.~DE-FG03-93ER40773 and by the Robert A. Welch Foundation under Grant
No.~A-1397.

\end{document}